\documentclass[prb, preprint]{revtex4-1}
\usepackage{graphicx}
\usepackage{amsmath}
\usepackage{amsfonts}
\usepackage{siunitx} 
\usepackage{bm}

\begin{document}
\title{Force on a moving object in an ideal quantum gas}
\author{Wittaya Kanchanapusakit}
\affiliation{Department of Physics, King Mongkut's University of Technology Thonburi, Pracha Uthit Road, Bangkok 10140, Thailand}
\author{Pattarapon Tanalikhit}
\affiliation{Department of Physics, Korea Advanced Institute of Science and Technology, Daejeon 34141, Republic of Korea}

\date{\today}
	
\begin{abstract}
		
We consider a heavy external object moving in an ideal gas of light particles. Collisions with the gas particles transfer momentum to the object, leading to a force that is proportional to the object's velocity but in the opposite direction. In an ideal classical gas at temperature $T$, the force acting on the object is proportional to $\sqrt{T}$. Quantum statistics causes a deviation from the $\sqrt{T}$-dependence and shows that the force scales with $T^2$ at low temperatures. At $T=0$, the force vanishes in a Bose gas but is finite in a Fermi gas. 

\end{abstract}
	
\maketitle

\section{Introduction}

An object moving in a fluid provides a platform to study hydrodynamics and kinetics of the medium material. The two effects are in the opposite regimes of the Knudsen number $\textrm{Kn} =\ell/R$, where $\ell$ is the mean free path of molecules in the fluid, and $R$ is the length scale of the object. In the hydrodynamic regime where $\textrm{Kn} \ll 1$ the fluid appears continuous to the object, and the principles of fluid mechanics apply. This leads to, for example, the well-known Stokes' formula for the force on a moving sphere, \cite{Stokes,Fluid_Mechanics} which can be extended to include the effect of finite values of Kn.\cite{Knudsen} This hydrodynamic force is proportional to $R$. In the opposite regime where $\textrm{Kn} \gg 1$, the medium particles, such as in a dilute gas, move freely and interact with the external object, resulting in a kinetic force.\cite{Kinetic_force,Drag_force1,Drag_force2,physical_kinetics} The magnitude of the kinetic force depends on the collision rate which scales with the cross section $\sim R^2$ of the object. Quantum mechanical effects enter via quantum scattering theory, giving an appropriate cross section for the scattering.\cite{drag_on_sphere} However, quantum distribution functions of the medium particles have yet to be taken into account.

We are interested in the motion of a heavy particle in a light gas. The study includes the mobility of an ion in a quantum fluid.\cite{Brownian_in_fermi,Clark,ion_in_helium4} A similar treatment will apply to polarons in a degenerate Fermi gas as well as in a Bose gas or a Bose condensate.\cite{object_in_fermi,object_in_bose} In this paper, we find the kinetic force from the known thermodynamic quantities of the gas. The model adopted consists of a massive impurity object embedded in an ideal quantum gas. Section II reviews Bose-Einstein and Fermi-Dirac functions. In section III, we derive a general expression for the kinetic force acting on the moving object. With a suitable quantum distribution function, the temperature dependence of the force is determined. Comparison is made between the force in the ideal quantum gas and that in the ideal classical gas. For simplicity, we let the Boltzmann constant $k_B$ equal to one and the Planck constant equal to $2\pi$ ($\hbar =1$) throughout.

\section{Bose-Einstein and Fermi-Dirac functions}

For a system of particles, each of which has mass $m$ and momentum $\bm{p}$, confined in a volume, we use Bose-Einstein or Fermi-Dirac distribution function in the form
\begin{equation}
	n_{\bm{p}} = \frac{1}{z^{-1}e^{p^2/2mT} +a} = \frac{1}{z^{-1}e^{x} +a},
\label{dist functions}
\end{equation} 
where $a = -1$ for bosons and $a=1$ for fermions. The dimensionless variable $z=e^{\mu/T}$ is called the fugacity, where $\mu$ is the chemical potential at temperature $T$, and we define $x=p^2/2mT$. The total number of the gas particles is given by $N=\sum_{\bm{p}}n_{\bm{p}}$. By considering the density of states in a volume $V$, the summation can be expressed as an integral over $sVd^3 p/(2\pi)^3=sV4\pi p^2dp/(2\pi)^3$, where the spin degeneracy factor $s$ is equal to 1 for spinless bosons and 2 for fermions. In terms of $x$-integral, the number density $n=N/V$ is given by
\begin{equation}
	n  = \frac{s(mT)^{3/2}}{\sqrt{2}\pi^2}\int_0^\infty \frac{\sqrt{x}dx}{z^{-1}e^{x} +a} = \frac{2s}{\sqrt{\pi}\lambda_T^3} \int_0^\infty \frac{\sqrt{x}dx}{z^{-1}e^{x} +a},
\label{n}
\end{equation}
where $\lambda_T = \sqrt{2\pi/mT}$ is the thermal wavelength. In our system of units ($k_B=1, \hbar=1)$, the mean thermal speed of the gas particles, in order of magnitude, is $\sqrt{T/m}$. Hence, $m\sqrt{T/m}=\sqrt{mT}$ has the dimension of momentum. Since the length has the dimension of reciprocal of momentum, $\lambda_T = \sqrt{2\pi/mT}$ (or $\lambda_T = \sqrt{2\pi\hbar^2/mk_B T}$ in SI units) has the dimension of length.
Integrals of the type $\int_0^\infty x^{r-1}dx/(z^{-1}e^x+a)$, $r>0$ are often encountered in Bose or Fermi systems. They can be evaluated:
\begin{equation}
	\int_{0}^{\infty} \frac{x^{r-1}dx}{z^{-1}e^x+a}=\Gamma(r)h_r(z),
	\label{h_functions}
\end{equation}
where $\Gamma(r)$ is the gamma function. The functions $h_r(z)$ are known as either the Bose-Einstein integral $g_r(z)$ or the Fermi-Dirac integral $f_r(z)$:\cite{Pathria, Dingle}
\begin{equation}
	h_r(z)= \begin{cases}
		g_r(z)      & a = -1, \\
		f_r(z)  	& a=1,
	\end{cases}
\label{h}
\end{equation}
where
\begin{equation}
	g_r(z)=\sum_{k=1}^{\infty} \dfrac{z^k}{k^r}, \qquad 	f_r(z)=\sum_{k=1}^{\infty} (-1)^{k-1} \dfrac{z^k}{k^r}.
\label{g and f}
\end{equation}
It follows from Eqs. (\ref{n}) and (\ref{h_functions}) that the number density can be expressed as
\begin{equation}
		n = \frac{s h_{3/2}(z)}{\lambda_T^3},
\label{number density}
\end{equation}
Quantum mechanical effects dominate at low temperatures when $\lambda_T$ is approximately equal to the interparticle distance, or when $n\lambda_T^3 \geq 1$.

\begin{figure}[t]	
		\centering
            \includegraphics[width=8.5 cm]{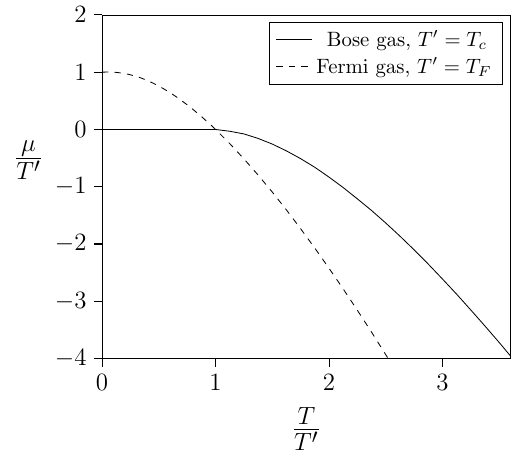}
		\caption{Chemical potential of an ideal Bose gas (solid line) and an ideal Fermi gas (dashed line).}
		\label{mu graph}
\end{figure}

Another important quantity is the chemical potential $\mu(T)$. In the Bose gas, there is a condensation temperature $T_c$ below which $\mu=0$ $(z=1)$. In the Fermi gas, the chemical potential is finite at $T=0$, and the quantity $\mu(0)$, also denoted by $T_F$, is called the Fermi energy. Equation (\ref{n}) or (\ref{number density}) can be used to plot $\mu(T)$, and the result is shown in Fig. \ref{mu graph}. At high temperatures (classical limit), one can see that $z = e^{\mu/T} \ll 1$. Equations (\ref{h}) and (\ref{g and f}) show that $h_r(z) \approx z$, $z \ll 1$, and it follows from Eq. (\ref{number density}) that the chemical potential for a given $n$ in the high-temperature limit is given by
\begin{equation}
    \mu  = T\ln\left( \frac{n\lambda_T^3}{s} \right),
\label{mu}
\end{equation}
where $n\lambda_T^3 \ll 1$.

\section{Kinetic force on moving object}

The consideration of the momentum transfer in this section follows Ref. \onlinecite{physical_kinetics} with an application of quantum statistics. The mass of the impurity object is taken to be large compared with that of the gas particle. The equilibrium distribution function of the particles is also assumed. 

\subsection{Momentum transfer and force}

\begin{figure}[h]	
		\centering
            \includegraphics[width=8.5 cm]{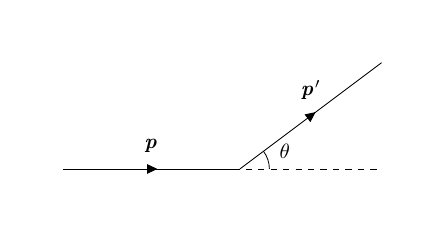}
		\caption{Change in momentum of a gas particle from $\bm{p}$ to $\bm{p'}$ by scattering through angle $\theta$.}
		\label{Mom_change}
\end{figure}

Consider a system of gas particles whose equilibrium distribution function $n_{\bm{p}}$ depends on the velocities $\bm{v}=\bm{p}/m$ of the particles. An impurity object is moving with a velocity $\bm{u}$ through the gas. The distribution function in the frame of the object is $n_{\bm{p}+m\bm{u}}$. Figure \ref{Mom_change} shows the scattering of a gas particle with momentum $\bm{p}$ off the object through an angle $\theta$. The momentum after the collision becomes $\bm{p'}$. The incoming particle flux is equal to $nv=n(p/m)$. This flux multiplied by the scattering cross section $d\sigma$ is equal to the number of particles scattered per unit time. Therefore, the differential force in the original direction $\alpha$ is given by
\begin{equation}
	dF^{\alpha}=(\bm{p}-\bm{p'})^\alpha n_{\bm{p}+m\bm{u}} \frac{p}{m}d\sigma.
\label{force_delta_p}
\end{equation}
For elastic collisions, it can be shown that the momentum transfer along the original direction is  $\bm{p}(1-\cos{\theta})$. Equation (\ref{force_delta_p}) becomes
\begin{equation}
	dF^\alpha = p^\alpha(1-\cos\theta) n_{\bm{p}+m\bm{u}} \frac{p}{m} d\sigma.
\label{differential_force}
\end{equation}
By summing over $\bm{p}$, one obtains the force acting on the external object:
\begin{equation}
		F^\alpha  =  \frac{1}{m}\sum_{\bm{p}}n_{\bm{p}+m\bm{u}}p p^\alpha \sigma_t(p),
		\label{force_sum}
\end{equation}
where
\begin{equation}
		\sigma_t(p)=\int(1-\cos\theta)d\sigma 
\label{transport_cross_section}
\end{equation}
is the transport cross section for the collisions. 

Assuming that $\bm{u}$ is much smaller than the speed of the gas particles, one can expand $n_{\bm{p}+m\bm{u}}=n_{\bm{p}}+m\bm{u}\cdot\partial n_{\bm{p}}/\partial \bm{p}$. In calculating the force in Eq. (\ref{force_sum}), the first term in the expansion is integrated to zero over all $\bm{p}$, leaving
\begin{equation}
	F^\alpha = \sum_{\bm{p}} p p^\alpha u^\beta\frac{\partial n_{\bm{p}}}{\partial p^\beta} \sigma_t(p).
\label{force}
\end{equation}
From the distribution function in Eq. (\ref{dist functions}), it follows that $\partial n_{\bm{p}} / \partial p^{\beta}=-(p^\beta/mT)n_{\bm{p}}(1-an_{\bm{p}})$. After averaging $\langle p^\alpha p^\beta \rangle = (p^2/3)\delta^{\alpha\beta}$ over the directions of $\bm{p}$, one obtains
\begin{equation}
	\bm{F} =-\frac{\bm{u}}{3mT}  \sum_{\bm{p}} p^3 n_{\bm{p}}(1 - an_{\bm{p}}) \sigma_t(p).
\label{drag quantum gas}
\end{equation}
The negative sign indicates that the force is in the opposite direction to the velocity of the object, hence a drag force. The factor $(1-an_{\bm{p}})$ depends on the relevant statistics. In the Fermi gas, the exclusion principle allows only the scattering that takes the gas particles into unoccupied states. This fact is included in the factor $(1-n_{\bm{p}}), a=1$. In quantum theory, the partial wave analysis can provide the appropriate scattering cross section.\cite{drag_on_sphere,Quantum_Mechanics}
	
To the first approximation, the transport cross section is taken to be a constant, denoted by $\sigma$. With the use of the identity
\begin{equation}
	\int_0^\infty x^{r-1} n_{\bm{p}}\left( 1-an_{\bm{p}}\right) dx =
	\Gamma(r) h_{r-1}(z),
\label{int bose function}
\end{equation}	
the force in Eq. (\ref{drag quantum gas}) can be calculated, yielding
\begin{equation}
	\bm{F} = -\frac{4s}{3\pi^2} \sigma m^2 T^2 h_2(z)\bm{u}.
\label{drag quantum gas after sum}
\end{equation}
To see the physical meaning of the above equation, let us consider the mean speed of the gas particles $\langle v \rangle = (1/n) \sum_{\bm{p}}(p/m)n_{\bm{p}}$ which is equal to
\begin{equation}
 \langle v \rangle = \sqrt{\frac{8T}{\pi m}} \frac{h_{2}(z)}{h_{3/2}(z)}.
\label{mean speed}
\end{equation}
Equations (\ref{number density}) and (\ref{mean speed}) are used to express Eq. (\ref{drag quantum gas after sum}) as
\begin{equation}
	\bm{F} = -\frac{4}{3} \sigma mn \langle v\rangle\bm{u}.
\label{drag quantum gas mean speed}
\end{equation}
The force is in the form $\bm{F}=-\gamma \bm{u}$, where $\gamma = 4\sigma m n \langle v \rangle /3$ is known as the drag coefficient. The quantity $\sigma n\left\langle v\right\rangle$ is interpreted as the collision rate. Equation (\ref{drag quantum gas mean speed}) has the same form as the classical force where the Maxwell-Boltzmann distribution is used.\cite{Kinetic_force,drag_on_sphere} In an ideal classical gas, one can find that $\langle v \rangle = \sqrt{8T/\pi m}$, and the force is proportional to $\sqrt{T}$. Apart from using the quantum scattering cross section, the quantum treatment enters via the use of Eq. (\ref{number density}) for the number density and  Eq. (\ref{mean speed}) for the mean speed of the gas particles.

In our model, the Knudsen number is $\ell/R \gg 1$. The kinetic theory of gas approximates $\ell \sim 1/n\sigma$, where $\sigma \sim R^2$. This gives the condition for the gas density $n \ll 1/R^3$. Hence, we have $1/\lambda_T^3 \leq n \ll 1/R^3$ for the quantum gas and $n \ll 1/\lambda_T^3 \ll 1/R^3$ for the classical gas. One can use $n \sim 1/d^3$, where $d$ is the mean interparticle distance, to compare the different relevant length scales.

\subsection{Ideal Bose gas}
\begin{figure}[h]	
		\centering
            \includegraphics[width=8.5 cm]{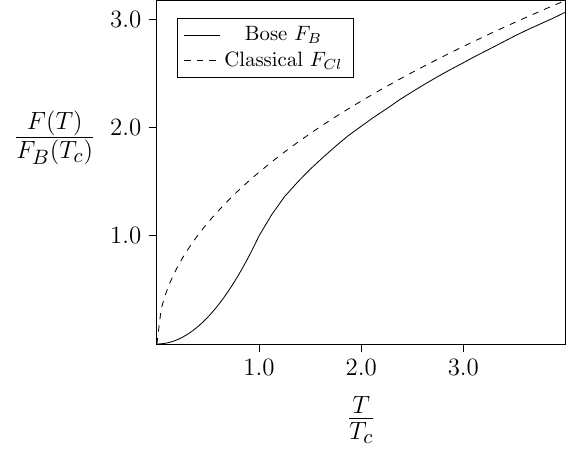}
		\caption{The kinetic force on an object moving in an ideal Bose gas. The dashed line represents $F_{Cl}(T)/F_B(T_c)$ where $F_{Cl}(T)$ is the force in the case of an ideal classical gas.}
		\label{Force Bose}
\end{figure}
In an ideal Bose gas, $h_2(z)$ is equal to $g_2(z)$. With $s=1$ for spinless bosons, if follows from Eq. (\ref{drag quantum gas after sum}) that the force on the object, denoted by $\bm{F}_B$, is 
\begin{equation}
	\bm{F}_B = -\frac{4}{3\pi^2} \sigma m^2 T^2 g_2(z)\bm{u}.
\label{drag bose}
\end{equation}
The temperature dependence of the force is contained in $T^2 g_2(z)$. The function $g_2(z)=\sum_{k=1}^{\infty}z^k/k^2$, $z=e^{\mu/T}$ can be evaluated at different temperatures by using $\mu(T)$ in Fig. \ref{mu graph}. The magnitude of the force as a function of temperature is shown in Fig. \ref{Force Bose}. For large $T$, the force approaches its classical value.

For $T\leq T_c$, we have $\mu=0$, causing $z=1$. When $z=1$, the function $g_r(1)$ is the Riemann zeta function $\zeta(r)$. Because of $g_2(1)=\zeta(2)=\pi^2/6$, Eq. (\ref{drag bose}) becomes
\begin{equation}
	\bm{F}_B = -\frac{2}{9} \sigma m^2 T^2\bm{u}, \quad T\leq T_c.
\label{drag quantum gas below Tc}
\end{equation}
The inflection point at $T=T_c$ on the solid line in Fig. \ref{Force Bose} is attributed to the formation of the condensate. One can use Eq. (\ref{drag quantum gas mean speed}) with the number density of the normal (non-condensate) particles $n \sim T^{3/2}, T\leq T_c$ and the mean speed $\langle v \rangle \sim \sqrt{T}$ to explain the $T^2$-dependence of the force for $T \leq T_c$. The force vanishes at $T=0$ where all the particles are in the condensate. Since the radius of the object is much smaller than the mean distance between the gas particles, the slowly moving object rarely experiences any collisions. At higher temperatures, there are more non-condensate particles whose collisions with the object contribute to the force. It is assumed that the collisions do not take the gas particles into or out of the condensate. There is a correction to the force in Eq. (\ref{drag quantum gas below Tc}) if the gas particles are allowed to change types via different scattering processes. 

\subsection{Ideal Fermi gas}

\begin{figure}[h]
	\centering
        \includegraphics[width=8.5 cm]{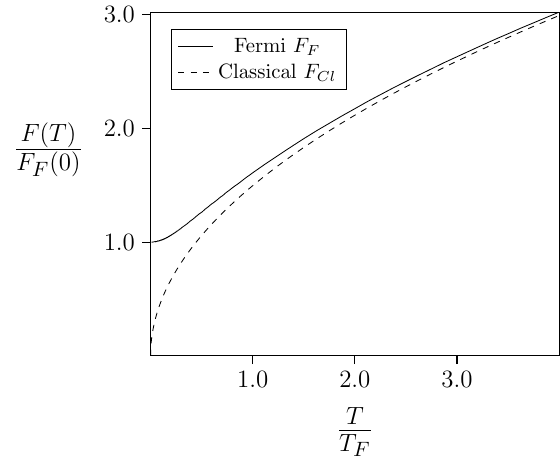}
	\caption{The kinetic force on an object moving in an ideal Fermi gas. The dashed line represents $F_{Cl}(T)/F_F(0)$ where $F_{Cl}(T)$ is the force in the case of an ideal classical gas.}
\label{Force Fermi}
\end{figure}

With $h_2(z)=f_2(z)$ and $s=2$, Eq. (\ref{drag quantum gas after sum}) gives the force, denoted by $\bm{F}_F$, exerted on a moving object in a Fermi gas:
\begin{equation}
	\bm{F}_F = -\frac{8}{3\pi^2} \sigma m^2 T^2 f_2(z)\bm{u}.
\label{drag fermi}
\end{equation}
By using the graph $\mu(T)$ in Fig. \ref{mu graph}, the function $f_2(z)=\sum_{k=1}^{\infty}(-1)^{k-1}z^k/k^2$, $z=e^{\mu/T}$ can be found at different temperatures. The temperature dependence of the force is shown in Fig. \ref{Force Fermi}. The force approaches its classical value for large $T$.

At low temperatures $(T \ll T_F)$, $z$ is large. We use expansion \cite{Pathria2}
\begin{equation}
	f_r(e^\xi)=\frac{\xi^r}{\Gamma(r+1)}\left[ 1+r(r-1)\frac{\pi^2}{6}\frac{1}{\xi^2}+...\right], 
	\label{f expand}
\end{equation}
together with Sommerfeld formula for chemical potential \cite{Blundell}
\begin{equation}
	\mu (T) = T_F\left(1-\frac{\pi^2}{12}\frac{T^2}{T^2_F} \right), \quad  T \ll T_F,
\label{sommerfeld}
\end{equation}
to approximate $f_2(z)$ in Eq. (\ref{drag fermi}) and derive
\begin{equation}
	\bm{F}_F = -\frac{4}{3\pi^2} \sigma m^2 T^2_F  \left(1+ \frac{\pi^2}{6}\frac{T^2}{T^2_F}\right) \bm{u},  \quad T\ll T_F.
\end{equation}
At $T=0$, the Fermi sphere of radius $p_F$, the Fermi momentum, is filled. The number density $n \sim p_F^3$ and the mean speed $\langle v \rangle \sim p_F$ lead to, from Eq. (\ref{drag quantum gas mean speed}), the fact that the magnitude of the force is scaled with $p_F^4$. By the definition of $T_F = p_F^2/2m$, the magnitude of the force is proportional to $T_F^2$. At low nonzero temperatures, those fermions with momenta $\delta p_F$ near the Fermi surface participate in the collisions and give rise to the force. The number of such fermions near this surface is proportional to $T^2$. 

\section{Conclusion}
	
The force exerted on a slowly moving object in an ideal quantum gas is studied in the regime of a large Knudsen number. The collisions with the gas particles are assumed to be elastic, and the corresponding transport cross section is taken to be constant. Such classical object in the quantum gas provides a semiclassical description of the kinetic force. At high temperatures, the distribution of the gas particles follows the Maxwell-Boltzmann type, and the force acting on the object is proportional to $\sqrt{T}$. At low temperatures, the quantum statistics causes the magnitude of the force to scale with $T^2$. Bosons in the zero-energy condensate rarely collide with the object because the size of the object is small compared with the mean spacing between the particles. Hence, the force in the Bose gas diminishes at $T=0$. In the Fermi gas, it is those fermions near the Fermi surface that scatter off the object and contribute to the force.

\section{Author declarations}

The authors have no conflicts to disclose.

\end{document}